\documentclass[aps,prd,12pt,preprintnumbers,nofootinbib,superscriptaddress]{revtex4}
\usepackage{graphicx}
\usepackage{mathrsfs}
\usepackage{amssymb,amsmath}
\def\beq{\begin{equation}}
\def\eeq{\end{equation}}
\def\sideremark#1{\ifvmode\leavevmode\fi\vadjust{\vbox to0pt{\vss
\hbox to 0pt{\hskip\hsize\hskip1em
\vbox{\hsize3cm\tiny\raggedright\pretolerance10000
 \noindent #1\hfill}\hss}\vbox to8pt{\vfil}\vss}}}

\newcommand{\bo}{\raise-1mm\hbox{\Large$\Box$}}

\newcommand{\f}[2]{\frac{#1}{#2}}

\newcommand{\w}{\omega}

\newcommand{\be}{\begin{equation}}
\newcommand{\ee}{\end{equation}}
\newcommand{\bea}{\begin{eqnarray}}
\newcommand{\eea}{\end{eqnarray}}
\newcommand{\bes}{\begin{subequations}}
\newcommand{\ees}{\end{subequations}}

\begin{document}

\title{Black Hole - Moving Mirror I: An Exact Correspondence}
\author{Paul R. Anderson}

\address{Department of Physics, Wake Forest University,\\
Winston-Salem, North Carolina  27109, USA\\
E-mail: anderson@wfu.edu}

\author{Michael R.R. Good}

\address{Physics Department, Nazarbayev University,\\
Astana, Republic of Kazakhstan\\
E-mail: michael.good@nu.edu.kz}

\author{Charles R. Evans }
\address{Department of Physics and Astronomy, University of North Carolina,\\
Chapel Hill, North Carolina 27599, USA\\
E-mail: evans@physics.unc.edu }

\begin{abstract}
An exact correspondence is shown between a new moving mirror trajectory in (1+1)D and a spacetime in (1+1)D
in which a black hole forms from the collapse of a null shell.  It is shown that the Bogolubov coefficients between
the ``in'' and ``out'' states are identical and the exact Bogolubov coefficients are displayed.  Generalization to the (3+1)D black hole case is discussed.
\end{abstract}


\maketitle

\vfill\eject

\vspace{0.5cm}
One motivation for studying the particle production from an accelerating mirror in (1+1)D is that if the non-uniformly accelerated trajectory is asymptotically null,
the particles are emitted at late times in a thermal distribution.\cite{Davies:1976hi, Davies:1977yv}  This is exactly what happens to the radiation emitted at
late times after a black hole forms from collapse.\cite{Hawking:1974sw}  This provides a well-known correspondence between accelerating mirrors and black holes.
An interesting question is whether there are mirror trajectories for which the entire history of the particle creation from its initial non-thermal phase
to its late time thermal distribution corresponds to the entire history of the particle creation from a spacetime in which a black hole forms from collapse.

Here we show a specific example in which this is the case.  The model for gravitational collapse consists of a collapsing shell with a null trajectory.  The
spacetime geometry inside the shell is flat while the geometry outside the shell is the usual Schwarzschild geometry.  The trajectory for the mirror is a simple modification
of one that was discovered in Ref.~\cite{thesis}.  The mirror begins at rest at $x = +\infty$ and accelerates in a monotonic fashion, asymptotically approaching $v = v_0$
with $v = t + x$.


\begin{figure}[h]
\centering
\includegraphics [trim=0cm 14cm 0cm 0cm,clip=true,totalheight=0.3\textheight]{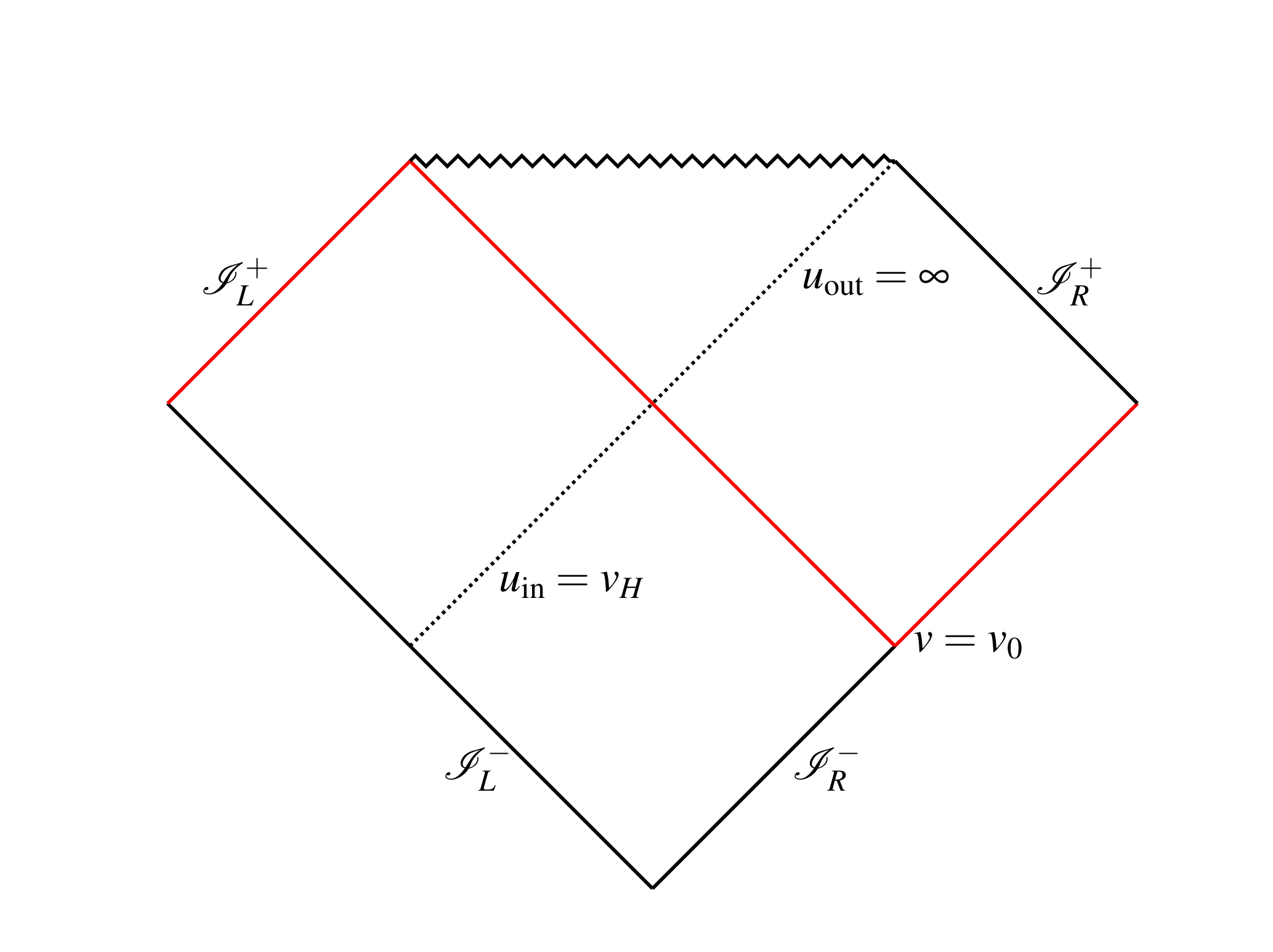}
\caption{ Penrose diagram for a 2D black hole that forms from the collapse of a null shell along the trajectory $v = v_0$.
The Cauchy surface used to compute the Bogolubov coefficients is shown in red.
Note that the horizon is the future light cone of the point $(u_{\rm in} = v_H \equiv v_0 - 4M \,, \, v = v_0$).}
\label{fig-bh_penrose}
\end{figure}

The Penrose diagram for the (1+1)D spacetime with a collapsing null shell is shown in Fig.~1.  We work with a massless minimally coupled scalar field.  Inside the shell
the space is flat.  The ``in'' modes are normalized on $\mathscr{I}^-$ in the same way the Minkowski modes are normalized there in flat space.
Thus everywhere inside the shell the modes are the same as they are in flat space with the result that
\bes
\bea \left(u^{\rm in}_{\w'} \right)_R &=& \frac{e^{-i \w' v}}{\sqrt{4 \pi \w'}}  \\
     \left(u^{\rm in}_{\w'} \right)_L &=& \frac{e^{-i \w' u_{\rm in}}}{\sqrt{4 \pi \w'}}
\eea \ees
with $v = t+r$ and $u_{\rm in} = t-r$.  There will also be a complete set of ``out'' modes with subsets which have three different late time behaviors.
 Some of the modes end on $\mathscr{I}^{+}_L$,  others go through the future horizon and end up at the singularity, and the rest end on $\mathscr{I}^{+}_R$.  We shall focus on the latter subset.  In the Schwarzschild region outside the shell these modes are
\be \left(u^{\mathscr{I}^+}_\w\right)_R = \frac{e^{-i \w u_{\rm out}}}{\sqrt{4 \pi \w}} \;.  \ee
Here $u_{\rm out} = t_s - r^*$, with $t_s$ the time in the usual Schwarzschild coordinates and $r^* \equiv r + 2 M \log[(r-2M)/(2M)] $.

To compute the number of particles produced which reach $\mathscr{I}^{+}_R$ we write
\be \left(u^{\mathscr{I}^+}_\w\right)_R = \int_0^\infty d \w' [ \alpha_{\w \w'} \left(u^{\rm in}_{\w'} \right)_L + \beta_{\w \w'} \left(u^{\rm in}_{\w'} \right)_L^* ] \label{future-modes} \ee
The Bogolubov coefficients can be obtained in the usual way using the scalar product~\cite{b-d-book}
\be  (\phi_1,\phi_2) = -i \int d \Sigma \sqrt{g_\Sigma} n^a [\phi_1 \nabla_a \phi_2^* - (\nabla_a \phi_1) \phi_2^* ]  \ee
with $\Sigma$ a Cauchy surface and $n^a$ the unit normal to that surface.  We can choose the surface that consists of $v = v_0$ plus the part of
 $\mathscr{I}^{-}_R$ with $v > v_0$ and $\mathscr{I}^{+}_L$, which is shown in red in the Penrose diagram in Fig.~1. The modes $\left(u^{\mathscr{I}^+}_\w\right)_R$ are nonzero only on the part of the Cauchy surface with $v = v_0$ which is outside the event horizon.
To do the matching we need to connect the inside and outside coordinates.  We do so by letting $r$ be continuous across the
surface.  Then~\cite{Fabbri:2005mw}
\be  u_{\rm out} = t_s - r^* = u_{\rm in} - 4 M \log \left[\frac{v_H - u_{\rm in}}{4 M} \right]   \ee
  where $v_H \equiv v_0 - 4M$.  After some calculation, the details of which will be given elsewhere~\cite{gae2}, we find
\bes
 \bea   \alpha_{\w \w'} &=& - \frac{2M \sqrt{\w \w'}}{\pi (\w-\w')} \, e^{-i(\w-\w') v_H} \, [- 4 M i (\w - \w')]^{-4 M \w i} \, \Gamma(4 M \w i)  \\
            \beta_{\w \w'} &=&  -\frac{2M \sqrt{\w \w'}}{\pi (\w +\w')} \, e^{-i(\w+\w') v_H} \, [- 4 M i (\w + \w')]^{-4 M \w i} \, \Gamma(4 M \w i) \;.
\eea  \label{alpha-beta-exact}
\ees

Next consider an accelerating mirror in (1+1)D flat space with a quantized massless scalar field which vanishes at the location of the mirror.  It is well known~\cite{Good:2013lca} that
the ``in'' modes are
\be  (u^{\rm in}_{\w'})_{\rm mirror} = \frac{1}{\sqrt{4 \pi \w'}} \left( e^{-i \w' v} - e^{-i \w' p(u)} \right)  \;, \ee
and the ``out'' modes which approach $\mathscr{I}^{+}_R$ are
\be (u^{\rm out}_{\w})_{\rm mirror} = \frac{1}{\sqrt{4 \pi \w}} \left(e^{-i \w f(v)}\, \theta(v_H-v) - e^{-i \w u} \right)  \;, \ee
where the ray tracing functions $p(u) $ and $f(v)$ are defined so that at the location of the mirror $p(u) = v$ and $f(v) = u$.  To compute the amount of particle production that occurs one can
expand $(u^{\rm out}_{\w})_{\rm mirror}$ in terms of $(u^{\rm in}_{\w'})_{\rm mirror}$ in the same way as was done for the spacetime with a null shell.
Computing the scalar product by integrating over the surface $\mathscr{I}^{-}$ gives expressions for  $\alpha_{\w \w'}$ and $\beta_{\w \w'}$.

A slight modification of the Omex trajectory in Ref.~\cite{thesis} gives
\be \label{trajectory} z(t) =  v_H -t - \f{W(2 e^{2 \kappa(v_H - t)})}{2\kappa}, \ee
with $W$ the Lambert W function and $\kappa$ and $v_H$ constants.  A transcendental inversion gives
\be  f(v) = v - \frac{1}{\kappa} \log [ \kappa (v_H - v) ]  \;. \ee
Again writing the ``out'' modes which are normalized on $\mathscr{I}^{+}_R$ and vanish on $\mathscr{I}^{+}_L$, in terms of the ``in''
modes which are normalized on $\mathscr{I}^-$, one can use the scalar product computed on $\mathscr{I}^-$ and obtain for $\kappa = (4M)^{-1}$
values for $\alpha_{\w \w'}$ and $\beta_{\w \w'}$ which are the same as those in~\eqref{alpha-beta-exact} for the spacetime with a collapsing null shell.

The (3+1)D version of a black hole that forms from the collapse of a null shell differs somewhat from the (1+1)D version although there are many similarities. The late time Hawking radiation
for this model has been discussed in~\cite{vilkovisky,Fabbri:2005mw}.  In (3+1)D there are scattering effects due to the spacetime geometry and the angular momentum (if it is nonzero) of the mode function.  The Penrose diagram is also somewhat different.  However, as will be shown elsewhere~\cite{gae2}, if scattering effects are ignored then the Bogolubov coefficients are given by the same expressions as in~\eqref{alpha-beta-exact} to within an overall, irrelevant, minus sign.

We have shown that in (1+1) D there is an exact one to one correspondence between the
 particle production that occurs for a massless minimally coupled scalar field when a black hole forms from gravitational collapse of a null shell and when a mirror in flat space has the trajectory~\eqref{trajectory}.  The time dependence of the particle production and its approach to a thermal distribution is given in the second paper of this series\cite{paper2}.

\section*{Acknowledgments}

PRA and CRE would like to thank Alessandro Fabbri and Amos Ori for helpful conversations.   This work was supported in part by the National
Science Foundation under Grant Nos. PHY-0856050, PHY-1308325, and PHY-1505875 to Wake Forest University, and PHY-1506182 to the University of North Carolina, Chapel Hill.


\begin{thebibliography}{0}
\bibitem{Davies:1976hi}
  P.~C.~W.~Davies, S.~A.~Fulling,
  Proc.\ Roy.\ Soc.\ Lond.\ A {\bf 348} (1976) 393.
	
\bibitem{Davies:1977yv}
  P.~C.~W.~Davies, S.~A.~Fulling,
  Proc.\ Roy.\ Soc.\ Lond.\ A {\bf 356}  (1977) 237.

	
\bibitem{Hawking:1974sw}
Stephen Hawking, 
  Commun. Math. Phys. \textbf{43} (1975) 199.
		
	\bibitem{thesis}
	M.~R.~R.~Good, 
    Ph.\ D.\ thesis, University of North Carolina at Chapel Hill, (2011).
			
\bibitem{Fabbri:2005mw}
  A.~Fabbri and J.~Navarro-Salas, {\it Modeling black hole evaporation} (Imperial College Press, London, UK, 2005).
	
\bibitem{gae2}
  M.~R.~R.~Good, P.~R.~Anderson and C.~R.~Evans, Manuscript in preparation.

\bibitem{Good:2013lca}
  M.~R.~R.~Good, P.~R.~Anderson and C.~R.~Evans,
  Phys.\ Rev.\ D {\bf 88}, 025023 (2013)
  [arXiv:1303.6756 [gr-qc]].

\bibitem{b-d-book}
See e.\ g.\  N.~D.~Birrell and P.~C.~W.~Davies,
{\it Quantum Fields in Curved Space},
Cambridge University Press (Cambridge, 1982).

\bibitem{vilkovisky} G.~Vilkovisky,
 in {\it Quantum theory of gravity; essays in honor of the 60th birthday of Bryce S. DeWitt},
 edited by  S.~M.~Christensen and Adam Hilger,
 (Bristol [Avon], 1984).

\bibitem{paper2} M.~R.~R.~Good, P.~R.~Anderson,  and C.~R.~Evans, 
to appear in the Proceedings of the 14th Marcel Grossmann Meeting, (2015).

	
\end{thebibliography}
\end{document}